\documentstyle[12pt,aasms4]{article}

\begin{document}

\slugcomment{MS 39279 --- May 25, 1999}
\lefthead{Furuya et al.}
\righthead{10 AU-Scale Jet-like flow in Class 0 Source S106 FIR}

\title{VLA Observations of $\bf H_2O$ Masers in the Class 0 Protostar S106 FIR:
Evidence for a 10 AU-Scale Accelerating Jet-like Flow}

\author{Ray S. FURUYA}
\affil{Department of Astronomical Science, 
The Graduate University for Advanced Studies,
Nobeyama Radio Observatory}
\authoraddr{Nobeyama 411, Minamimaki-mura, Nagano 384-1305, Japan}
\authoremail{(RSF) ray@nro.nao.ac.jp}

\author{Yoshimi KITAMURA}
\affil{Institute of Space and Astronautical Science}
\authoraddr{Yoshinodai, Sagamihara, Kanagawa 229, Japan}

\author{Masao SAITO}
\affil{Harvard-Smithsonian Center for Astrophysics}
\authoraddr{60 Garden St, Cambridge, MA 02138, USA}

\author{Ryohei KAWABE}
\affil{Nobeyama Radio Observatory}
\authoraddr{Nobeyama 411, Minamimaki-mura, Nagano 384-1305, Japan}

\and

\author{H. Alwyn WOOTTEN}
\affil{National Radio Astronomy Observatory}
\authoraddr{520 Edgemont Road, Charlottesville, VA 22903-3475, USA}

\begin{abstract}

We conducted VLA observations at $0.\arcsec 06$ resolution of 
the 22 GHz water masers toward the Class 0 source S106 FIR
($d=$ 600 pc; $15\arcsec$ west of S106-IRS4) on two epochs 
separated by $\sim$3 months.
Two compact clusters of the maser spots were found 
in the center of the submillimeter core of S106 FIR.
The separation of the clusters was $\sim$ 80 mas (48 AU) 
along $\rm P.A.=70^{\circ}$
and the size of each cluster was $\sim$20 mas$\times$10 mas.
The western cluster, which had three maser components,
was $\rm 8.0~km~s^{-1}$ blueshifted and the eastern cluster, 
which has a single component,
was $\rm 7.0~km~s^{-1}$ redshifted 
with respect to the ambient cloud velocity.
Each component was composed of a few spatially localized 
maser spots and was aligned on a line connecting the clusters.
We found relative proper motions of the components with
$\approx 30$ mas year$^{-1}$ (18 AU year$^{-1}$) along the line.
In addition, a series of single-dish observations show
that the maser components drifted with a radial accelerations of
$\rm \sim 1~km~s^{-1}~year^{-1}$.\par

These facts indicate that the masers could be excited by a 10
AU-scale jet-like accelerating flow ejected from 
an assumed protostar located between the two clusters.
The outflow size traced by the masers was 50 AU$\times$5 AU after
correction for an inclination angle of $\rm 10^{\circ}$ which 
was derived from the relative proper motions and radial velocities
of the maser components.
The three-dimensional outflow velocity ranged from 
40 to $\rm 70~km~s^{-1}$
assuming symmetric proper motions for the blue and red components.
Since no distinct CO  molecular outflows have been detected so far, 
we suggest that S106 FIR is an extremely young protostar observed
just after the onset of outflowing activity.\par

\end{abstract}


\section{INTRODUCTION}

The formation mechanism of jets and molecular outflows from young
stellar objects (YSOs) and their roles in the star formation process
are not well understood observationally or theoretically. 
One of the best observational approaches is to investigate very young
protostars which are in an evolutional stage just after the onset 
of the outflow activity.
Andr\'e, Ward-Thompson and Barsony\markcite{And93} (1993)
identified such protostars with sources which have not been detected 
in the near-infrared and which display a blackbody-like spectral energy
distribution (SED) which peaks in the submillimeter.
They proposed to classify these objects as 'Class 0' sources, 
being in a younger evolutional stage than that of 'Class I' sources
defined by Lada\markcite{Lad87} (1987). 
The cold SEDs of Class 0 sources suggest that the bulk of final stellar
mass has not yet been accumulated.
Class 0 sources are thought to be in their main accretion phase and 
they are known to have very powerful ``jet-like'' CO bipolar molecular 
outflows which are usually observed with typical sizes of $\sim 0.05-0.1$pc
(e.g., 
Bachiller et al.\markcite{Bac91} 1991;
Bachiller et al.\markcite{Bac96a} 1996;
Bachiller\markcite{Bac96b} 1996;
Yu and Chernin\markcite{Yu97} 1997).
In contrast, the CO outflows from Class I sources are
poorly collimated and much less powerful
(Bontemps et~al.\markcite{Bon96} 1996). 
It is believed that the acceleration and collimation of the outflows involve
magnetohydrodynamic processes in the vicinity of the central stars 
($r\lesssim 1~\rm AU$). 
For example, the outflows are thought to be centrifugally 
accelerated from magnetized Keplerian disks
(e.g., Uchida and Shibata\markcite{Uch85} 1985;
Kudoh and Shibata\markcite{Kud97} 1997; 
Pudritz et al.\markcite{Pud91} 1991) or thought to be accelerated
from the interaction regions between the stellar magnetospheres and 
the surrounding disks
(e.g., Shu et al.\markcite{Shu88} 1988;
Najita et~al.\markcite{Naj94} 1994).
Millimeter-interferometer observations with small arrays 
have extensively investigated the kinematics of 
gas around Class 0 sources.
Their spatial resolutions, however, were at most arcsecond scale 
(corresponding to a 100 AU scale in nearby star forming regions).
It is essential to attain higher spatial resolution -- down to 
subarcsecond-scales -- in order to understand 
outflows in the vicinity of Class 0 sources.\par

Interferometric imaging of the $\rm H_2$O maser line at 22 GHz 
with the VLA and VLBA provides an excellent tool for studying 
subarcsecond structure in 
protostellar jets very close to protostars.
It has been suggested that the maser lines in low-mass 
star forming regions are associated with outflows, because the
luminosities of the masers correlate with mass-loss rates derived from CO 
molecular observations
(Felli et al.\markcite{Fel92} 1992) and with the luminosities of 
6 cm free-free continuum emission from ionized, 
collimated outflows
(Wilking et al.\markcite{Wil94} 1994;
Meehan et al.\markcite{Mee98} 1998).
Clear evidence for that was obtained by high spatial resolution studies 
with the VLA-A array (e.g., Chernin\markcite{Che95} 1995) 
and with the VLBA
(e.g., Claussen et al.\markcite{Cla98} 1998).
These observations revealed that the masers originate 
behind the shocked gas very close to the central protostars.
Interferometric observations of the maser lines have three advantages.
First, we can attain high angular resolutions ( $\sim$1-100 mas ) 
and high relative positional accuracies
(typically $\sim 10\%$ of angular resolutions).
Second, the maser lines are free from extinction 
even in deeply embedded sources.
Third, we can attain high velocity resolution 
($\lesssim 0.1~\rm km~s^{-1}$) with radio spectroscopy.
However, such studies have just begun toward the Class 0/I sources.\par

We have found that Class 0 sources provide the best targets
for studying gas kinematics very close to protostars through
observations of the $\rm H_2$O maser lines. 
Masers 
have been detected for 18 of 30 Class 0 sources 
(detection rate is $\approx 60\%$) in our ongoing 
multi-epoch survey toward Class 0/I sources 
with the Nobeyama 45m-telescope
(Furuya et al.\markcite{Fur98} 1998).
On the other hand, we measured a detection rate of only 
$\approx$ 10\% for Class I sources.
This estimate is based upon data compiled
from our survey and previous ones
(Wouterloot and Walmsley\markcite{Wou86} 1986; 
Wilking and Claussen\markcite{Wil87} 1987;  
Cessaroni et al.\markcite{Ces88} 1988; 
Comoretto et al.\markcite{Com90} 1990;
Terebey et al.\markcite{Ter92} 1992; 
Felli et al.\markcite{Fel92} 1992; 
Palagi et al.\markcite{Pal93} 1993; 
Xiang and Turner\markcite{Xia95} 1995; 
Claussen et al.\markcite{Cla96} 1996),
although sensitivities and angular and velocity
resolutions differ.
We interpreted this result as follows.
Since Class 0 sources possess a large amount of circumstellar gas,
the gas shocked by outflow can easily reach the physical conditions
necessary to excite the maser emission 
($10^6\rm cm^{-3}\lesssim {\it n}(\rm H_2)\lesssim 10^9\rm cm^{-3}$
with $T\gtrsim a~few~ hundred$ K
; Hollenbach\markcite{Hol97} 1997, or
$n(\rm H_2)\gtrsim 10^9\rm cm^{-3}$
with $T\gtrsim 300$ K
; Elitzur et al.\markcite{Eli89} 1989 ).
In addition to providing appropriate conditions near the outflow origin,
edge-on disks could also enhance the detectability of maser emission by 
maximizing the path length of velocity coherent gas
( Elmegreen and Morris\markcite{Elm79} 1979). 
In this case, 
we could investigate kinematics of protostellar disks using the
maser probe as shown by Fiebig et al.\markcite{Fie96} (1996)
in IRAS 00338$+$6312.\par

In this paper, we have studied the 22 GHz $\rm H_2$O maser emission 
toward the Class 0 source S106 FIR.
A compact, bipolar HI$\hspace{-0.03truecm}$I region 
lies near S106 FIR.
A distance to S106 molecular cloud is 600 pc
(Eiroa, Els$\ddot{\rm a}$sser and Lahulla\markcite{Eir79} 1979;
Staude et al.\markcite{Sta82} 1982).
The exciting star S106-IRS4
(other names are S106-IRS3, S106-IR and S106 PS)
whose spectral type is O8V-B0V (Gehrz et al.\markcite{Geh82} 1982),
is located $15\arcsec$ east of S106 FIR.
S106 FIR is a Class 0 source isolated from this bipolar
HI$\hspace{-0.03truecm}$I region.
A dense core with a diameter of 9000 AU was observed 
around S106 FIR in 450, 800 and $1100\mu$m continuum emission 
using the JCMT 
(Richer et al.\markcite{Ric93} 1993), 
while no emission at 20 $\mu$m was detected 
(Gehrz et al.\markcite{Geh82} 1982). The SED of S106 FIR suggests 
that this source is a Class 0 source. 
Analysis of the SED implies that the luminosity of S106 FIR is
in the range of $\approx 24$ to $10^3 L_{\sun}$
(Richer et al.\markcite{Ric93} 1993). 
It should be noted that a distinct CO molecular outflow has not been 
detected in S106 FIR, making it unique among all Class 0 sources 
searched
(Hayashi et al.\markcite{Hay93} 1993; Bachiller 1996\markcite{Bac96};
Furuya et al. 1999\markcite{Fur99}), 
although there is a possibility that 
wing emission from the S106FIR outflow may be hidden by
broad line emission associated with 
the HI$\hspace{-0.03truecm}$I region.
The peak of the submillimeter core
(see Fig. 4 of Richer et al.\markcite{Ric93} 1993) 
corresponds to the position of the H$_2$O masers.
The maser emission was first detected by 
Stutzki et al.\markcite{Stu82} (1982) and subsequent observations were
made by Kawabe\markcite{Kaw87} (1987) using the 
Nobeyama Millimeter Array (NMA) in 1987.
He found that the masers 
have a doubly-peaked spectrum with a velocity coverage of 
15 km $\rm s^{-1}$
and that all the maser spots are concentrated within an area of radius 
$0.\arcsec 3$ (180 AU).
At this position, no free-free radio continuum emission has been detected 
(Felli et al.\markcite{Fel84} 1984; Bally et al.\markcite{Bal83} 1983; 
Hoare et al.\markcite{Hoa94} 1994).
The presence of the $\rm H_2O$ masers, dense molecular cloud core and 
the cold SED indicate that S106 FIR may harbor a protostar.
In particular, the absence of distinct CO molecular outflows suggests 
that this protostar is in an extremely early state of star formation.\par

\section{OBSERVATIONS}

Deep integration synthesis observations of the $\rm H_2O$ maser line 
($\rm 6_{16}-5_{23}$; $\nu_{\rm rest}=$ 22235.077 MHz) toward S106 FIR 
were performed with the 
NRAO\footnote{The National Radio Astronomy Observatory (NRAO) is operated 
by Associated Universities, Inc., under cooperative agreement with the 
National Science Foundation} Very Large Array (VLA) in its A and BnA 
configurations on 1996 October 15 and 1997 January 23, respectively. 
The FWHM of the primary beam was $2.\arcmin 0$ 
and the system noise temperatures 
were around 170  K. We observed in the single IF-mode using 128
channels with a total bandwidth of 3.125~MHz. The velocity
range covered was $\rm 42.1 ~km~s^{-1}$ 
with a velocity resolution of $\rm 0.33~km~s^{-1}$ at 
22.235 GHz (Table \ref{tbl:VLAobs}).
The phase tracking center in the A configuration observations was
located at the position reported in the previous NMA measurements 
(${\rm R.A.}_{1950}=20^h~25^m~32.^s45$, 
${\rm Dec.}_{1950}=37^{\circ}~12\arcmin ~50.\arcsec 95$; 
Kawabe\markcite{Kaw87} 1987).
The center in the BnA configuration observations was chosen to be
at the strongest maser spot in the A configuration observations.
The integration time per visibility was 8 seconds, and the field of
view (FOV) was limited to be $37\arcsec$ by a coherency level of 97 \% 
for the longest baseline.\par

\placetable{tbl:VLAobs}

The continuum source $2005+403$ was used as a phase and amplitude
calibrator, and was  
observed for 2 minutes during each 15-minute observing cycle. 
For the first epoch, $2005+403$ was also used as a bandpass calibrator
by integrating over the run. For the second epoch, 3C454.3 was observed
as a bandpass calibrator using 3-minute integration 
at the beginning and end of the observations.
The flux density scale for the first epoch was established using the
continuum source $0404+768$ which was not resolved even in the 
A configuration.
The flux density scale for the second epoch was set by
the continuum source 3C286.
According to the VLA List of Calibrator Sources, the flux densities of 
$0404+768$ and 3C286 at $\lambda =$ 1.3 cm were 
1.0 Jy and 2.5 Jy, 
respectively, and the accuracy is expected to be better than 30\%.
We made maps of the central regions of 128 mas $\times$ 128 mas area
in the FOV using task IMAGR in the AIPS package with a cell size of 0.5 mas.
The resulting synthesized beam sizes were 63 mas $\times$ 61 mas at 
P.A. = $85^{\circ}$ and 250 mas 
$\times$ 80 mas at P.A. = $89^{\circ}$ for the first and second epochs, 
respectively.\par

We obtained the absolute positions of the strongest maser spots at the
velocity channels of
$\rm V_{LSR}=~6.4 ~km~s^{-1}$ 
( ${\rm R.A.}_{1950}=20^h~25^m~32.^s533\pm 0.^s00015$, 
${\rm Dec.}_{1950}=37^{\circ}~12\arcmin ~50.950\arcsec\pm 0.002\arcsec$)
and 
$\rm ~-13.7 ~km~s^{-1}$ 
( ${\rm R.A.}_{1950}=20^h~25^m~32.^s450\pm 0.^s0006$, 
${\rm Dec.}_{1950}=37^{\circ}~12\arcmin ~50.788\arcsec\pm 0.0025\arcsec$)
for the first and second epochs, respectively, before a self-calibration 
procedure.
The absolute position accuracies of the strongest spots
were estimated to be about 
$\sigma_{\rm R.A.,Dec}\approx$ 4 mas for the first epoch and
$\sigma_{\rm R.A.}\approx$ 14 mas, $\sigma_{\rm Dec}\approx$ 5 mas for the
second epoch, respectively,
taking account of the typical baseline error of the array 
($|\Delta\bf b|\rm \approx 1~cm$)
and the angular separation of the phase calibrator from the source
($|\Delta\bf s\rm|=4.9^{\circ}$).
These excellent absolute position accuracies result from 
the use of the close phase calibrator whose coordinates are known 
with a very high positional accuracy ($\lesssim 2$ mas).\par

In order to obtain the relative positions of the remaining 
maser spots to the strongest spots at individual channels and 
to improve dynamic range of images, 
we have employed a self-calibration
procedure using only the phase of the strongest maser spots.
The typical rms noise levels in the channel maps after the 
self-calibration were improved to 10.5 mJy beam$^{-1}$ 
from 28.5 mJy beam$^{-1}$ 
and 9.4 mJy beam$^{-1}$ from 22.5 mJy beam$^{-1}$ 
for the first and second epochs, respectively.
The relative positional accuracies were determined mainly by the
signal-to-noise ratio (S/N ratio), because the 
bandpass calibration was done within an accuracy of a few degrees
across the channels, and because the self-calibration worked very well 
and the residual systematic errors were very small.
The relative positional error of a point source convolved with a 
synthesized Gaussian beam is given by 
$0.45~\theta$/(S/N ratio) where $\theta$ is the FWHM
of the beam (Reid et al.\markcite{Rei88a} 1988a;
$\theta=\sqrt{\theta_{\rm maj}\times\theta_{\rm min}}$).
For the first epoch, 
we use only the data with S/N ratio higher than 7, resulting
in a maximum positional error of 5.6 mas (3.4 AU).
For the second epoch data, we take the S/N ratio higher
than 16 which gives the same maximum positional error
as that of the first epoch.
These relative positional errors are comparable to those of
the absolute ones except for $\sigma_{\rm R.A.}\approx$ 14 mas of the
second epoch.\par

\placetable{tbl:H2Oobs}               

We also monitored
the masers irregularly in order to
investigate the velocity drifts of the maser peaks.
The maser emission was observed in 1987 May using the NMA (Kawabe 1987). 
Single-dish observations were made with the 45m-telescope of 
Nobeyama Radio Observatory 
(NRO)\footnote{Nobeyama Radio Observatory (NRO) is a branch of National 
Astronomical Observatory, an interuniversity research institute
operated by the Ministry of Education, Science, and Culture of Japan}
on 1996 May 5 and 28.
Also 5-minute snapshot observations with the NRAO, 
Very Long Baseline Array (VLBA) were conducted on 1996 June 5 
as a VLBI Space Observatory Program pre-launch survey 
(Migenes et al.\markcite{Mig98} 1998).
The observational parameters are summarized in Table \ref{tbl:H2Oobs}.\par

\section{RESULTS}

\subsection{Time Variation of the Overall $\bf H_2O$ Maser Spectrum}
\label{sub:Varia}

Figure \ref{fig:H2Osp} shows the $\rm H_2O$ maser spectra obtained by
the monitoring observations. 
The intensity scale is shown in Jy unit except for the VLBA
snap shot observations because of the uncertainties in the
flux calibration.
The spectra commonly show three major peaks on the blueshifted side 
and a single peak on the redshifted side with respect to the ambient 
cloud velocity ($V_{\rm sys}=\rm -1.1~km~s^{-1}$) measured from 
$\rm H^{13}CO^+$ $J=1-0$ observations with the NMA 
(Furuya et al.\markcite{Fur98} 1998)
and 
$\rm NH_3$ $(J,K)=(2,2)$ observations with the VLA
(Mangum and Wootten\markcite{Man94} 1994).
Distinct maser emission cannot be seen around the ambient cloud 
velocity.
The total radial velocity range of the masers is always from 
$V_{\rm LSR}\approx\rm~-15$ to $\rm 12 ~km~s^{-1}$.
The blue peaks are commonly seen in the velocity range from 
$V_{\rm LSR}\approx -15$ to $-2$ $\rm km~s^{-1}$.
The red peak is seen from 
$V_{\rm LSR}\approx +2$ to $+12~\rm km~s^{-1}$, 
although the peak velocity changed from 
$V_{\rm LSR}=+9$ to $\rm +6~km~s^{-1}$ in the third quarter of 1996.\par

\placefigure{fig:H2Osp}

We found velocity drifts of the spectral peaks in the spectra of 
Figure \ref{fig:H2Osp}.
The velocities of the three distinct blue peaks drifted 
together toward the blue, particularly from May 1996 till January
1997. The blue peaks at
$V_{\rm LSR} = -7.4,~-9.4~\rm and ~-12.1$ $\rm km~s^{-1}$ 
drifted to 
$V_{\rm LSR} = -7.8,~-9.8~\rm and ~-13.7$ $\rm km~s^{-1}$, respectively,
during the two epochs of VLA mapping.
In addition, we found similar peak velocity drifts 
on the red side at
$V_{\rm LSR} = +9~\rm km~s^{-1}$
from May 1996 until June 1996 and at
$V_{\rm LSR} = +6~\rm km~s^{-1}$
during the VLA observations.
The red peak at
$V_{\rm LSR} = 9.2~\rm km~s^{-1}$ drifted to 
$V_{\rm LSR} = 9.5~\rm km~s^{-1}$ in about one month and the peak at
$V_{\rm LSR} = 6.4~\rm km~s^{-1}$ drifted to 
$V_{\rm LSR} = 6.7~\rm km~s^{-1}$ in about three months.
Here, we fit Gaussian shapes to the spectra in order to
determine the peak velocities.
These peak velocity drifts can be clearly seen in Figure \ref{fig:EpochVel} 
which shows the peak velocities as a function of the 
observing date. \par

The important point which we found from Figure \ref{fig:EpochVel} 
is that individual peaks seem
to have their own constant accelerations along the line of sight.
For the blue peaks, 
the accelerations of the velocity drifts are estimated to be
$-2.0,~-1.1~\rm and -1.1$ 
$\rm km~s^{-1}year^{-1}$ for the peaks 1, 2 and 3, respectively; 
and their mean is 
$\rm -1.4~km~s^{-1}year^{-1}$.
For the red peaks, the peak at $V_{\rm LSR}\approx\rm 9~km~s^{-1}$ 
showed a velocity drift with an acceleration of 
$\rm +3.9~km~s^{-1}year^{-1}$ on May 1996.
The peak at $V_{\rm LSR}\approx\rm 6~km~s^{-1}$ 
showed a velocity drift with an acceleration of 
$\rm +1.2~km~s^{-1}year^{-1}$ during the VLA
observations.
These velocity drifts are summarized in Table \ref{tbl:velocity}.\par

\placefigure{fig:EpochVel}
\placetable{tbl:velocity}

\subsection{Spatial Structures of Maser Spots}

Figure \ref{fig:map} shows the relative peak positions of the 
$\rm H_2O$ maser spots in channels of $0.66~\rm km~s^{-1}$ 
resolution, achieved by combining together
two adjacent individual channels, 
for the two epochs of our VLA observations.
We plotted maser spots with S/N ratios higher 
than 7 and 16 for the first and
and second epochs, respectively.
Map origins plotted by double-circles show the positions of the 
phase referencing maser spots.
A spectrum of the masers integrated over the 
128 mas$\times$128 mas area is also shown with 
a velocity resolution of $0.33~\rm km~s^{-1}$ 
at the upper right corner of each panel. 
Even at the highest resolution of $0.33~\rm km~s^{-1}$,
significant differences cannot be seen in the position and
position-velocity maps when compared with 
Figures \ref{fig:map} and  \ref{fig:pv}
at a resolution of $0.66~\rm km~s^{-1}$.
On the other hand, $0.99~\rm km~s^{-1}$ resolution exceeds
the typical line width of the spectral peaks and causes 
a serious degradation on the definition 
of maser components (see $\S$3.4).
Thus, we employ a resolution of $0.66~\rm km~s^{-1}$ in the maps
to avoid the confusion caused by many points.\par

\placefigure{fig:map}

We found that two clusters of maser spots are located in the
center of the submillimeter core around S106 FIR
and that no maser spots 
are found outside the 
128 mas $\times$ 128 mas central area.
This result is consistent with the previous NMA results
( Kawabe\markcite{Kaw87} 1987).
The apparent separation between the intensity-weighted central
positions of the two clusters is
76.2 mas ( 45.7 AU) at $\rm P.A.=72^{\circ}$ and 87.5 mas ( 52.5 AU) 
at $\rm P.A.=70^{\circ}$ for the first and second epochs,
respectively. Each cluster is elongated along a line connecting 
the two clusters.\par

\subsection{Velocity Structures of Maser Spots}
\label{sub:VelSt}

Two clusters of the maser spots were found toward the Class 0 
source S106 FIR.
The western and eastern clusters of maser spots are 
$\rm \sim 5-13~\rm km~s^{-1}$ blueshifted and 
$\sim 6-\rm 10~\rm km~s^{-1}$
redshifted compared to the ambient cloud velocity, respectively.
Hereafter, we call them blue and red clusters.
The velocity coverages of the blue and red clusters are 
approximately symmetric with respect to the ambient cloud
velocity.\par

In order to investigate internal velocity structures within 
each cluster, we show position-velocity maps along two orthogonal 
lines in Figure \ref{fig:pv}.
The top and bottom panel in Figure \ref{fig:pv} are
the position-velocity maps along and perpendicular 
to a line connecting the two clusters.
This line is not only parallel to the elongation direction of the
clusters but also parallel to proper motions 
which we will describe in $\S 3.5$.
We determined the line by $\chi^2$-fitting and its 
position angle was measured to be $72^{\circ}\pm1^{\circ}$ for 
October 1996 and $70^{\circ}\pm 2^{\circ}$
for January 1997.
We adopt a mean of $\rm P.A.=71^{\circ}$.
In Figure \ref{fig:pv}, filled- and open-circles indicate 
the maser spots for the first and second epochs, respectively, 
and the vertical scales of the position-velocity diagrams are 
the relative radial velocities with respect 
to the ambient cloud velocity.\par

\placefigure{fig:pv}

As is obvious from the top panel of Figure \ref{fig:pv},
there exist systematic velocity gradients within each cluster
along the line at $\rm P.A.=71^{\circ}$, particularly
in the blue cluster.
The absolute value of velocity increases with increasing 
the distance from the middle of the two clusters.
Mean velocity gradients over the two epochs were 
$\rm -0.74~km~s^{-1}~AU^{-1}$ for the blue cluster and
$\rm -0.36~km~s^{-1}~AU^{-1}$ for the red, determined 
by $\chi^2$-fitting.
The velocity gradient for the blue cluster is twice as steep as
that of the red cluster.
The bottom panel of Figure \ref{fig:pv} shows the 
position-velocity diagram along the line at P.A.$=-19^{\circ}$.
All of the maser spots are distributed in a strip of 14 mas
(8.4 AU) width centered at the P.A.$=71^{\circ}$ line.
No clear velocity gradient can be seen along the line at
P.A.$=-19^{\circ}$ in each cluster.\par

\subsection{Identification of Maser Components}

In this subsection, we will identify groups of spatially 
localized maser spots corresponding to the major peaks in the 
spectra at each epoch. Then we will identify
the maser components across the two epochs.\par

In order to identify maser components in each epoch, 
we use the VLA images with
velocity resolution of $0.66~\rm km~s^{-1}$
(Figure \ref{fig:map}) and the spectra of the masers
(Figures \ref{fig:H2Osp} and \ref{fig:map}).
In the spectra of the masers,
we identified three distinct velocity peaks
($V_{\rm LSR}\approx -7,~-9$ and $ -12~\rm km~s^{-1}$)
on the blue side and a single velocity peak 
($V_{\rm LSR}\approx 6.0\rm~km~s^{-1}$) on the red side.
Here, we call the peaks at $V_{\rm LSR}\approx -7~\rm km~s^{-1}$ as
A1 and B1 on the first and second epochs, respectively.
Similarly, we 
label the peaks at $V_{\rm LSR}\approx -9$ and 
$-12~\rm km~s^{-1}$ as shown in Tables \ref{tbl:position96}
and \ref{tbl:position97}.
In the upper right corner of each panel of Figure \ref{fig:map},
the solid bars above the velocity axis show the radial velocity 
ranges of the three blue peaks and one red peak which we will 
identify as maser components.
In Figure \ref{fig:velmap}, we show maps of the relative positions
of maser spots by displaying the velocity ranges of the 
spectral peaks in pseudocolors.
In the upper right corner of each panel, we show the definition of
the velocity ranges of the peaks.
We found that the maser spots marked by individual colors
seem to be spatially localized.
Hence, the individual spectral peaks can be identified with these
spatial components.
We show the relative positions of the components in 
Tables \ref{tbl:position96} and \ref{tbl:position97}.\par

\placefigure{fig:velmap}

Next, we will show correspondence of the maser components 
between the two epochs.
The blue A1, A2 and A3 components on the first epoch are 
considered to be identical to the blue B1, B2 and B3 
components on the second epoch, respectively, from the following 
kinematical arguments.
First, each pair shows a similar velocity range over the two epochs.
Second, the peak velocity drifts over $\sim$250 days 
in Figure \ref{fig:H2Osp} strongly suggest the presence of three
kinematical components on the blueshifted side
(i.e., blue 1, 2 and 3), which move with their 
own constant accelerations (Figure \ref{fig:EpochVel}).
In addition, the red component has a positive acceleration
similar to the absolute values of the acceleration of the blue
components,
although the red component was observed only over the two epochs.
Third, the blue components show similar relative proper 
motions with respect to the red one, as will be
shown in $\S$3.5.

\placetable{tbl:position96}
\placetable{tbl:position97}

\subsection{Relative Proper Motions of Maser Components}
\label{sub:Prop}

We found an increase in the separations of about 10 mas over 
the two epochs between the red and three blue components in 
Figure \ref{fig:velmap}.
This fact means that there exist proper motions along the 
line connecting the two clusters.
The absolute positions of the phase referenced maser spots 
in the two epochs were used to bring self-calibrated relative 
position maps onto an absolute position basis.  This procedure 
complicates proper motion estimation.
Because typical absolute positional errors are
$\sigma'_{\rm R.A.,Dec}\approx$ 7 mas for the first epoch 
and
$\sigma'_{\rm R.A.}\approx$ 15 mas, 
$\sigma'_{\rm Dec}\approx$ 8 mas for the
second epoch, respectively, and are comparable to
the proper motions, the proper motions are imprecisely determined.\par

Nevertheless, we can derive relative proper motions from comparison of
the relative position maps whose map origins are at the positions of 
the red spot at $V_{\rm LSR}=\rm 6.4~km~s^{-1}$ for the first epoch and at
the red spot at $V_{\rm LSR}=\rm 6.7~km~s^{-1}$ for the second.
This superposition is based on our interpretation that
the two red spots are physically identical as described in $\S 3.4$.
We show a magnified H$_2$O maser spot map of the blue cluster for 
the two epochs in Figure \ref{fig:bothmap}.
Dashed-lines with labels denote the identified maser components.
We now find the relative proper motions of the components
over the two epochs.\par

\placefigure{fig:bothmap}

The distances between the red component and the three 
blue components with labels of 1, 2 and 3 
increased from 71.7, 73.2 and 83.7 mas to 84.1, 82.1 and 96.3 mas, 
over the two epochs, respectively.
The corresponding proper motions are 
$12.4\pm 5.4$, $8.9\pm 4.3$ and $12.6\pm 3.9$ mas for the
blue components 1, 2 and 3, respectively.
The proper motion derived for the component 3 was above the $3\sigma$
level, although the other two were at the $2\sigma$ level.
The directions of the motions are almost 
parallel to the line connecting the two clusters.
Table \ref{tbl:proper} summarizes the changes of the separations and
the relative proper motions.
The intensity weighted mean of the proper motions of 
the blue components relative to the red one is
$\rm 10.9\pm 4.2 ~mas$ ($\rm 6.5 \pm 2.5 AU$) per 100 days.
This can be converted to a relative transverse velocity of
$\rm 113\pm 44~km~s^{-1}$ along the plane of the sky.\par

\placetable{tbl:proper}

\section{DISCUSSION}

We will discuss three possible models for the origin of 
the two maser clusters:
(1) masers associate with a compact jet-like flow, 
(2) masers originate near each star of a binary system, and
(3) masers are generated in tangential parts of a rotating disk.
We conclude that the jet-like flow model will explain our results well.\par

\subsection{Compact Jet-like Flow Model}
\subsubsection{Evidence for Compact Outflow and Its Physical Properties}

The elongation of the spatial distributions of the maser spots,
the presence of the blue and red cluster of maser spots and 
the relative proper motions of the maser components along the
elongation strongly suggest
that the maser emission is excited
in an expanding jet-like outflow.
We assume that the driving source of the outflow, a protostar,
is located at the middle of the line connecting the two clusters.
This hypothesis is supported by the fact that the spectral profiles
of the masers have blue- and redshifted symmetric peaks with 
respect to the ambient cloud velocity which is thought to be the same as
the systemic velocity of the central protostar.\par

We will discuss three-dimensional geometry of the jet-like flow
based on our results.
First, we can derive an inclination angle of the outflow out of the 
plane of the sky from the proper motions and radial velocities 
of the maser components.
The inclination angle $i$ comes from
\begin{equation}
\tan i =\frac{|V_{\rm mean}-V_{\rm sys}|}{\case{1}{2}\mu}
\label{eqn:inclination}
\end{equation}
where $V_{\rm mean}$ is a mean peak velocity over the two epochs 
and $\mu$ is the relative proper motion for each blue component.
We obtained $i=5.7^{\circ}, 10.3^{\circ}$ and $10.2^{\circ}$ for the three 
blueshifted components, respectively (see Table \ref{tbl:proper}); 
we adopt an intensity weighted mean of $i=9^{\circ}$.
So, the outflow axis of S106 FIR is nearly in the plane of the sky.\par

Next, we can estimate an outflow velocity range using both the radial 
velocities and the relative proper motions.
Since symmetric proper motions for the blue and red components
are suggested by the symmetric maser spectra with respect to the
ambient cloud velocity,
the outflow velocity is estimated to be one half of the transverse
velocity described in $\S 3.5$ and ranges from 
45 to $65 \rm~km~s^{-1}$, corrected for the
inclination angle (see Table \ref{tbl:proper}).
This derived velocity range is comparable to that reported by
Claussen et al. (1998)\markcite{Cla98} who measured the proper 
motions of the H$_2$O masers in the Class I source IRAS 05413$-$0104.
The escape velocity from the protostar of S106 FIR ranges 
from 10.3 to 20.6 km~s$^{-1}$ at a distance of 25 AU from the 
protostar, taking the stellar mass to be 1.5 - 6 $M_{\sun}$ 
deduced from the observed bolometric luminosity range 
($24L_{\sun}\lesssim L_{\rm bol} \lesssim 1000L_{\sun}$).
Hence, the maser spots could not be gravitationally bound 
to the star: this is consistent with the outflow model.\par

Furthermore, we estimate the acceleration of the outflow from the 
radial velocity drifts of the maser components; 
we assume that the motions of the maser components are
limited to lie along the outflow axis.
The resulting acceleration ranges from 4 to 15 $\rm ~km~s^{-1}year^{-1}$ 
corrected for the inclination angle.
We summarize properties of the outflow in Table \ref{tbl:outflow}.\par

\placetable{tbl:outflow}

\subsubsection{Formation Mechanism of the Jet-like Flow}

We will discuss two acceleration mechanisms for the jet-like flow.
The masers could be accelerated by the stellar wind from 
the assumed protostar, or by stellar radiation.
We conclude that the former mechanism seems most plausible,
because the radiation pressure appears insufficient
to accelerate the jet-like flow up to the acceleration 
range estimated in $\S$4.1.1.\par

First, we will discuss the mechanism whereby the masers are 
accelerated by the stellar wind which is commonly thought to 
accelerate the water masers 
(e.g., Genzel et al.\markcite{Gen81} 1981).
Here, we consider each identified maser component as a spherical 
maser cloudlet with a radius of $a \approx 5$ mas (3AU),
which is a typical size of the identified components described
in $\S$3.4,
for simplicity.
Since the acceleration, $\alpha_{\rm W}$, is caused by a pressure 
gradient over the cloudlet, we have
$\alpha_{\rm W}=\frac{1}{\rho}\cdot\frac{\Delta P}{a}\sim
\frac{1}{\rho}\cdot\frac{P_{\rm W}}{a}$,
where $\rho$ is a mass density of the cloudlet and
$P_{\rm W}$ is a pressure from the stellar wind.
The mass density of the maser cloudlet, $\rho$ is 
$n_{\rm H_2}m_{\rm H_2}$,
where $n_{\rm H_2}$ is the number density of the cloudlet and
$m_{\rm H_2}$ is the weight of molecular hydrogen.
The wind pressure, $P_{\rm W}$ at a distance $r$ from the protostar
is given by
$\frac{\dot{M}_{\rm w}v_{\rm w}}{4\pi r^2}$, where  
$\dot{M}_{\rm w}$ and $v_{\rm w}$ are the mass loss rate 
(typically $\sim 10^{-7}M_{\sun}\rm year^{-1}$)
and wind velocity 
(typically $\sim 200~{\rm km~s}^{-1}$)
of protostellar jets, respectively
(e.g., Edwards, Ray and Mundt\markcite{Edw94} 1994).
Thus, we obtain the acceleration from stellar winds as follows
\begin{eqnarray}
\alpha_{\rm W} &\sim& \frac{1}{n_{\rm H_2}m_{\rm H_2}}\cdot\frac{1}{a}\cdot
\frac{\dot{M}_{\rm w}v_{\rm w}}{4\pi r^2} \nonumber\\
         &\sim& 7.5~{\rm km~s^{-1}~yr^{-1}}
\left(\frac{\dot{M}_{\rm w}}{10^{-7}M_{\sun}yr^{-1}}\right)
\left(\frac{v_{\rm w}}{200~{\rm km~s}^{-1}}\right)
\left(\frac{25~{\rm AU}}{r}\right)^2
\left(\frac{3~{\rm AU}}{a}\right)
\left(\frac{10^8{\rm ~cm^{-3}}}{n_{\rm H_2}}\right) \nonumber\\
 & &
\end{eqnarray}
This value seems to be roughly consistent with the acceleration range
of $4\sim 15~\rm km~s^{-1}~year^{-1}$ estimated in $\S 4.1.1$.
Here, we assumed a typical molecular hydrogen density of
$10^8\rm ~cm^{-3}$, necessary to produce maser emission.
If we consider smaller cloudlets, down to maser spot sizes of 1 AU
(e.g., Reid and Moran 1988b\markcite{Red88b}), 
$\alpha_{\rm W}$ becomes larger.
However this does not have a serious influence on our discussion.
Hence, it is probable that the jet-like flow traced by the
masers was mainly accelerated by the stellar wind pressure.\par

Next, we will discuss the acceleration, $\alpha_R$, due to
stellar radiation pressure. The acceleration from radiation 
pressure is given by $\frac{\kappa L_{\rm bol}}{4\pi cr^2}$.
If we take the bolometric luminosity, $L_{\rm bol}$ to range from 
24 to 1000$L_{\sun}$ and the mass opacity coefficient, 
$\kappa$ to be $230 \rm ~cm^2~g^{-1}$
(a typical value in interstellar medium), 
then the acceleration from the radiation pressure
ranges from 0.13 to $5.2~\rm km~s^{-1}~year^{-1}$ 
at $r=25$ AU from the star.
Here, we assume that the maser cloudlets are optically thin.
If we take an optically thick maser cloudlet, $\alpha_R$ becomes
smaller compared with the optically thin case.
This estimated range is smaller than the inclination-corrected
acceleration of $\sim 4-15~\rm km~s^{-1}~year^{-1}$.
Hence, it seems unlikely that stellar radiation pressure 
accelerates the jet-like flow.\par

\subsection{Other Models for the Origin of the Masers }

In this subsection, we discuss the remaining two models
for the origin of the masers.
First, we will discuss a model in which the masers originate near 
each star of a protobinary system. Recent infrared and optical 
observations of pre-main sequence (PMS) stars revealed that the 
frequency of binary systems in PMS stars is comparable to that 
in main sequence stars 
($\gtrsim 50\%$; e.g., Mathieu 1994\markcite{Mat94}: 
Ghez et al. 1997\markcite{Ghe97}: 
Kohler and Leinert 1998\markcite{Koh98}: 
Brandner and Kohler 1998\markcite{Bra98}).
This fact suggests that binarity must be common among protostars.\par

In the binary model, the separation of the two maser clusters can be 
taken as the projected binary separation 
($2a\arcmin=\rm 48~AU$).
This separation is near the median separation of binaries
($2a\arcmin =$30 AU from Duquennoy and Mayer 1991).
The orbital plane of the binary system is suggested to be
perpendicular 
to the plane of the sky because the proper motions were parallel to the line 
connecting the two clusters. This is a suitable configuration for the 
maser amplification. 
We estimate a rotational velocity of the binary stars in the
orbit as
$V_{\rm bin}=\sqrt{(V_{\rm LSR}-V_{\rm sys})^2+(\case{1}{2}\mu)^2}
\approx 60~\rm km~s^{-1}$.
A binary phase $\phi$ measured from the zero 
projected separation can be estimated from
$\tan\phi=|V_{\rm LSR}-V_{\rm sys}|/(\frac{1}{2}\mu)$ 
for the individual maser components. We adopt a mean value of 
$\phi=+10^{\circ}$. \par

The binary model has a serious drawback that the system is not
in a gravitationally bound state.
For the binary system to be gravitationally bound, the following
large enclosed mass is required.
\begin{equation}
M_{\rm enclosed}(r\leq R_{\rm maser})\sim 
560M_{\sun}\left(\frac{a}{140~{\rm AU}}\right)
\left(\frac{V_{\rm bin}}{60~\rm km~s^{-1}}\right)^2
\end{equation}
Here, $a$ is the half of the binary separation given by 
$a=a\arcmin/\sin\phi\approx 140$ AU.
This enclosed mass is equivalent to a reduced mass of the system.
However, the mass is considerably larger than the
reduced mass of the binary system expected from the bolometric
luminosity range. 
Therefore, we reject the protobinary model.\par

Next, we discuss a model of a protostellar disk: 
a single star is located between the clusters and 
the masers are excited in tangential parts of a rotating disk around
the star. Elmegreen and Morris\markcite{Elm79} (1979) originally studied 
such a model in an ideal case: 
if there exists enough column density of velocity 
coherent gas along the line of sight in an edge-on rotating disk, 
then $\rm H_2O$ masers can be emitted in tangential
parts of the disk. The resulting maser spectrum is expected to have a 
symmetric double-peaked shape with respect to the systemic velocity. 
The terminal velocity of the spectral peaks corresponds 
to the rotational velocities of the maser emitting regions of the disk.
Water masers originating in disks were found, 
for the first time, towards the Class I protostar 
IRAS 00338$+$6312 with the VLA
(Fiebig et al.\markcite{Fie96} 1996).
The maser profile defined a symmetric double-peaked
shape with respect to the ambient cloud velocity.
The velocity structure of the maser spots suggests that
the $\rm H_2O$ masers are associated with an 
infalling, rotating disk-like envelope whose radius, 
traced by the masers, ranges from 10 to 30 AU.\par

The disk model is not plausible in case of S106 FIR, although the
H$_2$O maser spectra in the source have a symmetric double-peaked
shape.
In S106 FIR, the disk plane is suggested to be 
nearly edge-on as mentioned in the binary model, and
we can estimate an enclosed mass of the disk and star 
in the same way as we did for the binary model.
The estimated enclosed mass is large, on the order of 100$M_{\sun}$,
which cannot be reconciled with the bolometric luminosity range.
Therefore, we reject the disk model.\par

\subsection{Future Works to Reveal the Nature of S106 FIR}

The presence of the $\rm H_2O$ masers and the cold SED suggests that 
the star formation has just begun in this source.
We stress that no well-developed outflows have been 
reported in S106 FIR (Hayashi et al 1993 and Bachiller 1996).
In addition, all of the Class 0 sources except S106 FIR are 
known to possess distinct CO  molecular outflows.
These facts suggest that S106 FIR is at
an evolutionary state just after the onset of outflow activity and
the formation of well-developed CO molecular outflows usually
observed. Do these masers represent a very early state in the 
development of an outflow from a protostar ? \par

We have searched for bipolar molecular outflows using the 
lines of CO {\it J}=1-0 and $\rm HCO^+$ {\it J}=1-0 with the 
NMA and the NRO 45m-telescope, respectively. 
However, no large-scale molecular outflow has been found
(Furuya et al.\markcite{Fur99} 1999).
On the other hand, $\rm H^{13}CO^+$ {\it J}=1-0 observations with the
NMA and $\rm NH_3$ ({\it J,K})=(1,1) and (2,2) observations with the
VLA showed the presence of a dense molecular cloud core around S106
FIR with a mass of $\sim 1.0-1.5M_{\sun}$ and a size of 9000 AU.
This dense core is expected to supply molecular gas to a large-scale
outflow through interaction with a stellar wind or the compact
jet-like outflow.\par

Our work is being continued as further studies of the proper motions
of the  $\rm H_2O$ masers with the VLBA.
The detailed structure and proper motions of the
maser spots will be discussed in a forthcoming paper together with
a search for large-scale CO outflow with the NMA.
A search for the exact position of the central 
star(s) by free-free continuum observations will also be attempted.
We propose that the central star would be located at the middle of the
two clusters 
( ${\rm R.A.}_{1950}=20^h~25^m~32.^s49$, 
${\rm Dec.}_{1950}=37^{\circ}~12\arcmin ~50.\arcsec 87$).
\par

\section{SUMMARY}

High angular resolution VLA observations of the 22 GHz water maser line 
were conducted toward the Class 0 source S106 FIR.
The main findings of this paper are as follows.\par

\begin{enumerate}

\item{We found two clusters of the $\rm H_2O$ maser spots
    with a separation of 50 AU, located at the center of the
    submillimeter core around the Class 0 protostar S106 FIR. The sizes of
    the clusters are 50 AU$\times$5 AU. The western and eastern
    clusters are $\rm 8.0~km~s^{-1}$ blueshifted and $\rm
    7.0~km~s^{-1}$ redshifted, respectively, 
    with respect to the ambient cloud velocity.
}

\item{Three maser components were identified 
    in the blue and a single one in the red clusters, respectively.
    These components are composed of spatially localized maser
    spots within a 10 mas (6 AU) area, and correspond to the major peaks in
    the overall spectrum of the $\rm H_2O$ maser line.
}

\item{We determined both the relative proper motions and peak velocity
    drifts of the maser components. The proper motions are along the line 
    connecting the two clusters and are $\sim 30$ mas year$^{-1}$.
    All the components showed similar peak velocity drifts with 
    radial accelerations of $\sim 1~\rm km~s^{-1}year^{-1}$.
}

\item{We conclude from our observations that the 
    masers are associated with a highly collimated compact
    accelerating jet-like flow which could originate from an assumed 
    protostar located at the middle of the line connecting the two clusters. 
}

\item{It is possible that S106 FIR is an extremely young protostar in
    an evolutionary state just after the onset of outflow activity,
    because of the absence of distinct CO molecular outflows.
}

\end{enumerate}

\acknowledgments

The authors gratefully acknowledge an anonymous referee for her very 
careful reading of the manuscript.
Mark J. Claussen and Kevin B. Marvel provided fruitful suggestions as 
collaborators of our ongoing VLBA study.
R. F. acknowledges the kind valuable help by V. Migenes and K. Shibata, 
during the course of data reduction process with the AIPS. 
R. F. also appreciates useful discussion and encouragements with 
M. Tsuboi, T. Omodaka, T. Nakano, S. Kameno, S. S. Hayashi, 
T. Kudoh  and H. Shinnaga.
In addition, authors would like to thank the staff of NRAO - Socorro and 
the NRO. M. S. is supported by Smithsonian Postdoctoral Fellowship program.\par






\begin{figure}
\caption{
Spectra of the $\rm H_2O$ maser emission toward S106 FIR observed with
the NMA (1987 May 8), the Nobeyama 45m-telescope (1996 May 5 and 28),
the VLBA (1996 June 5) and the VLA (1996 October 15 and 1997 January 23).
The intensity scales are shown in Jy unit except for the VLBA observations.
The ambient cloud velocity ($V_{\rm sys}=\rm -1.1~km~s^{-1}$),
measured from 
$\rm H^{13}CO^+$ $J=1-0$ and 
$\rm NH_3$ $(J,K)=(2,2)$ observations, is also shown. 
Arrows drawn on the spectra show the drifts of the
major peaks ( see in text).
}
\label{fig:H2Osp}
\end{figure}

\begin{figure}
\caption{
Time variations of the velocities of the major peaks of the $\rm H_2O$ 
maser spectrum in S106 FIR on the redshifted ({\it upper panel}) and
on the blueshifted ({\it lower panel}) sides.
The peak velocities are determined by applying Gaussian profiles
to individual peaks.
}
\label{fig:EpochVel}
\end{figure}

\begin{figure}
\caption{
Peak position maps of the $\rm H_2O$ maser spots shown with a velocity 
resolution of $\rm 0.66~km~s^{-1}$ with S/N ratio higher than 7 for 
(a) 1996 October 15 observations and higher than 16 for
(b) 1997 January 23.
The map origins shown by double-circles are defined by the positions of
the phase referencing maser spots at the
$V_{\rm LSR}=6.4~\rm km~s^{-1}$ component on 1996 October and 
at the $V_{\rm LSR}=-13.7~\rm km~s^{-1}$ component on 1997 January.
The spectrum of the masers integrated over the mapping region is 
also shown at the upper right corner of each
panel with a velocity resolution of $\rm 0.33~km~s^{-1}$. 
Solid bars above the velocity axis in the spectra indicate the ranges 
of the identified velocity components.
}
\label{fig:map}
\end{figure}

\begin{figure}
\caption{
Upper and lower panels: superposed position-velocity maps of the maser spots on 
1996 October 15 observations (filled circles) and on 1997 January 23 
(open circles).
The upper and lower panels show the position - velocity maps along the
line at P.A.$=71^{\circ}$ and P.A.$=-19^{\circ}$, respectively.
In the ordinate of the position-velocity maps, the ambient cloud velocity 
($V_{\rm sys}=-1.1~\rm km~s^{-1}$) is subtracted from the radial
velocities of the masers spots.
Center panel: superposed maps of the maser spots over the two
epochs. The two orthogonal lines indicate the lines of P.A.$=71^{\circ}$ and 
P.A.$=-19^{\circ}$, along which we made the position-velocity maps.
}
\label{fig:pv}
\end{figure}

\begin{figure}
\caption{
Peak position maps of identified maser components shown by pseudocolor.
The definitions of the components are indicated in the upper right 
corner of each panel. The map origins are the position of the 
$\rm 6.4~km~s^{-1}$ spot for the first epoch, and the position of the 
$\rm 6.7~km~s^{-1}$ spot for the second epoch.
The phase referencing maser spots in the maps are shown by white-circles.
}
\label{fig:velmap}
\end{figure}

\begin{figure}
\caption{
Superposed map of the maser spots at the two epochs for the blue
cluster. 
The positional accuracies are the same as in Figure 3, and the 
map origins are the same as in Figure 5.
Filled and open circles represent 1996 October and 1997 January 
observations, respectively.
Dashed lines with labels indicate the identified maser components.
}
\label{fig:bothmap}
\end{figure}

\setcounter{table}{0}
\begin{table}[hbtp]
\newcommand{\lw}[1]{\smash{\lower2.ex\hbox{#1}}}
\setcounter{table}{0}
\caption{\bf VLA Mapping Observations of $\bf H_2O$ Masers toward S106 FIR}
\label{tbl:VLAobs}
\bigskip
\begin{tabular}{lcccccc}
\hline\hline
Epoch & Config. & Integ. Time & \multicolumn{2}{c}{Synthesized
Beam} & Vel. Coverage & Vel. Resolution \\
\cline{4-5}
& & & Size & P.A. \\
& & (hours) & (mas) & (deg) & (km s$^{-1}$) & (km s$^{-1}$) \\
\hline \\
1996 Oct 15&A & 3 & $63\times61$ &85 &42.1 &0.3 \\
1997 Jan 23&BnA & 2 & $250\times80$ &89 &42.1 &0.3 \\
\hline
\end{tabular}
\end{table}

\begin{table}[hbtp]
\newcommand{\lw}[1]{\smash{\lower2.ex\hbox{#1}}}
\begin{center}
\caption{\bf Summary of $\bf H_2$O Maser Observations toward S106 FIR}
\bigskip
\label{tbl:H2Oobs}
\begin{tabular}{ccccccc}
\hline\hline
& & & \multicolumn{2}{c}{Beam Size} & & \\
\cline{4-5}
Date & & Telescope & Primary & Synthesized  & Vel. Resolution & Sensitivity\tablenotemark{a} \\
          & &                  &    ($\arcsec$)       &  (mas)      & ($\rm km~s^{-1}$) & (mJy) \\ \hline
1987 May ~8 & & NMA            & 330             &  5000       & 0.26              & $\sim$100  \\
1996 May ~5 & & NRO 45m        & 74  & $\cdot\cdot\cdot$       & 0.50              & 180  \\
1996 May 28 & & NRO 45m        & 74  & $\cdot\cdot\cdot$       & 0.50              & 240  \\
1996 Jun ~5 & & VLBA\tablenotemark{b} & 120  & $\approx$ 30\tablenotemark{c}  & 0.21  & 220   \\
1996 Oct 15 & & VLA-A          & 120 & 63  $\times$ 61                     & 0.33  & 10    \\
1997 Jan 23 & & VLA-BnA        & 120 & 250 $\times$ 79                     & 0.33  & 10    \\ \hline

\end{tabular}
\end{center}
\tablenotetext{a}{Rms noise level per velocity channel. }
\tablenotetext{b}{Performed as a pre-launch survey of VLBI Space Observatory 
Program (VSOP). }
\tablenotetext{c}{The size is estimated from the fringe spacing of the 
shortest baseline, between Los Alamos and Pie Town in the VLBA.}

\end{table}


\begin{table}[hbtp]
\newcommand{\lw}[1]{\smash{\lower2.ex\hbox{#1}}}
\begin{center}
\caption{\bf Radial Velocity Drifts of $\bf H_2O$ Maser Peaks in S106 FIR}
\label{tbl:velocity}
\bigskip
\begin{tabular}{lccc}
\hline\hline
\lw{Maser Component} & \multicolumn{2}{c}{Radial Velocity}  & \lw{Radial Acceleration} \\
\cline{2-3}
       & 15 Oct 1996   &   23 Jan 1997 &                              \\
       & (km s$^{-1}$) & (km s$^{-1}$) & (km s$^{-1}$ yr$^{-1}$) \\ \hline
Red    &          6.4  &     6.7       & $+1.2$                       \\
Blue 1 &       $-7.4$  &  $-7.8$       & $-2.0$                       \\
Blue 2 &       $-9.4$  &  $-9.8$       & $-1.1$                       \\
Blue 3 &       $-12.1$ & $-13.7$       & $-1.1$                       \\
Blue (mean) &  $\cdot\cdot\cdot$ & $\cdot\cdot\cdot$ & $-1.4$\tablenotemark{a}      \\
\hline
\end{tabular} 
\end{center}

\tablenotetext{a}{Intensity weighted mean}

\end{table}

\begin{table}[hbtp]
\newcommand{\lw}[1]{\smash{\lower2.ex\hbox{#1}}}
\begin{flushleft}
\caption{\bf Relative Positions of $\bf H_2O$ Maser Components in S106
  FIR on the First Epoch (1996 Oct. 15)}
\bigskip
\label{tbl:position96}
\begin{tabular}{cccccccc}
\hline\hline
\multicolumn{2}{c}{Maser Components}        & & \multicolumn{4}{c}{Position Offset\tablenotemark{a}} 
                                                                                   & Total Flux Density \\ 
                        &                   & &               & &                & &      \\
\cline{1-2}\cline{4-7}
Peak Velocity & \lw{Name}& & $\Delta$ R.A. & & $\Delta$ Dec.  & &      \\
($\rm km~s^{-1}$)       &                   & & (mas)         & & (mas)          & & (Jy) \\
                        &                   & &               & &                & &     \\  \hline
                        &                   & &               & &                & &     \\
   6.4\tablenotemark{b} & $\cdot\cdot\cdot$ & & 0.0           & & 0.0            & & 1.9 \\
 $-7.4$                 & A1                & & $68.9\pm 1.6$ & & $-19.7\pm 1.6$ & & 0.3 \\
 $-9.4$                 & A2                & & $70.5\pm 0.4$ & & $-19.7\pm 0.4$ & & 1.1 \\
$-12.1$                 & A3                & & $81.3\pm 0.4$ & & $-19.8\pm 0.4$ & & 1.3 \\ \hline
\end{tabular}
\end{flushleft}
\tablenotetext{a}{Intensity weighted mean position}
\tablenotetext{b}{Map origin}
\end{table}

\begin{table}[hbtp]
\newcommand{\lw}[1]{\smash{\lower2.ex\hbox{#1}}}
\begin{flushleft}
\caption{\bf Relative Positions of $\bf H_2O$ Maser Components in S106
  FIR on the Second Epoch (1997 Jan. 23)}
\bigskip
\label{tbl:position97}
\begin{tabular}{cccccccc}
\hline\hline
\multicolumn{2}{c}{Maser Components}        & & \multicolumn{4}{c}{Position Offset\tablenotemark{a}} 
                                                                                   & Total Flux Density \\ 
                        &                   & &               & &                & &      \\
\cline{1-2}\cline{4-7}
Peak Velocity  & \lw{Name}& & $\Delta$ R.A. & &
$\Delta$ Dec.  & &      \\
($\rm km~s^{-1}$)        &                   & & (mas)          & & (mas)          & & (Jy) \\
                         &                   & &                & &                & &      \\  \hline
                         &                   & &                & &                & &      \\
   6.7                   & $\cdot\cdot\cdot$ & & $-93.8\pm 2.8$ & & $22.0\pm 2.7$  & & 1.1 \\
 $-7.8$                  & B1                & & $-12.3\pm 1.6$ & & $1.09\pm 1.0$  & & 1.6 \\
 $-9.8$                  & B2                & & $-1.7\pm 3.4$  & & $-13.8\pm 2.6$ & & 2.2 \\
$-13.7\tablenotemark{b}$ & B3                & & 0.0            & & 0.0            & & 4.0 \\ \hline
\end{tabular}
\end{flushleft}
\tablenotetext{a}{Intensity weighted mean position}
\tablenotetext{b}{Map origin}

\end{table}


\begin{table}[hbtp]
\newcommand{\lw}[1]{\smash{\lower2.ex\hbox{#1}}}
\begin{flushleft}
\caption{\bf Relative Proper Motions of $\bf H_2$O Maser Components in S106 FIR}
\bigskip
\label{tbl:proper}
\begin{tabular}{cccccccc}
                           & &               &              &
                           &                   &                         & \\ \hline\hline
Name                       & & \multicolumn{2}{c}{Distance from Red Comp.} 
                                                            & $\mu$\tablenotemark{a}
                                                                            & $V_{\rm mean}$    & $i$\tablenotemark{b}  & $V_{\rm 3D}$\tablenotemark{c} \\
\cline{3-4}
                           & & 15 Oct 1996   & 23 Jan 1997  &            
                                                                            &                   &                       &                          \\
                           & & (mas)         & (mas)        & (mas)         & ($\rm km~s^{-1}$) & (degree)              & ($\rm km~s^{-1}$)  \\
                           & &               &              &
                           &                   &
                           &                     \\ \hline
Blue 1 (A1$\rightarrow$B1) & & 71.7$\pm$2.3  & 84.1$\pm$4.9 & 12.4$\pm$5.4  &  $-7.6$           &  5.7$^{+3.6}_{-1.9}$  & 64$^{+1.2}_{-5.0}$  \\
Blue 2 (A2$\rightarrow$B2) & & 73.2$\pm$0.6  & 82.1$\pm$4.3 &  8.9$\pm$4.3  &  $-9.6$           & 10.3$^{+11.2}_{-0.6}$ & 46$^{+1.0}_{-4.1}$  \\
Blue 3 (A3$\rightarrow$B3) & & 83.7$\pm$0.6  & 96.3$\pm$3.9 & 12.6$\pm$3.9  & $-12.9$           & 10.2$^{+11.5}_{-0.3}$ & 65$^{+1.1}_{-5.5}$  \\ \hline
\end{tabular}
\end{flushleft}
\tablenotetext{a}{Relative proper motions measured from the red component.}
\tablenotetext{b}{Inclination angle defined by
  $\arctan\left({\frac{\mid V_{\rm mean}-V_{\rm sys}\mid}{\mu /2}}\right)$,
where $V_{\rm sys}=~-1.1~\rm km~s^{-1}$.}
\tablenotetext{c}{Three-dimensional velocity obtained assuming symmetric proper motions for the blue
and red components.}

\end{table}


\begin{table}[hbtp]
\newcommand{\lw}[1]{\smash{\lower2.ex\hbox{#1}}}
\begin{flushleft}
\caption{\bf Physical Properties of Compact Jet-like Flow in S106 FIR }
\label{tbl:outflow}
%
\begin{tabular}{ccc}
                                                                  & &
                                                                  \\ \hline\hline
Size~ (AU)                                                        & & 50$\times$5 \\
P.A.~ (degrees)                                                    & & 70 \\
Inclination~ (degrees)                                             & & 9 \\
Outflow Velocity\tablenotemark{a}~ (km $\rm s^{-1}$)                               & & 45$\sim$65 \\ 
Outflow Acceleration\tablenotemark{b}~~(km $\rm s^{-1}~year^{-1}$) & & 4$\sim$15\\ \hline
\end{tabular}
\end{flushleft}
\tablenotetext{a}{Traced by the masers.}
\tablenotetext{b}{Acceleration, traced by the masers, corrected for the inclination angle of
  $9^{\circ}$ assuming that the motions of the maser components are
  along the outflow axis.}

\end{table}


\end{document}